\documentstyle[prl,aps,epsf,twocolumn]{revtex}
\begin{document}
\draft
\twocolumn[
\hsize\textwidth\columnwidth\hsize\csname @twocolumnfalse\endcsname


\title{ COEXISTENCE OF THE CRITICAL SLOWING DOWN AND GLASSY FREEZING 
        IN RELAXOR FERROELECTRICS}

\author{B. E. Vugmeister and H. Rabitz}
\address{Department of Chemistry, Princeton University, Princeton, NJ
08544}

\maketitle

\widetext
\begin{abstract}
We have developed a dynamical model for the dielectric response in
relaxor ferroelectrics
 which explicitly  takes into account  the coexistence of the critical 
slowing down and glassy freezing. 
The application of the model to the experiment in
PbMg$_{1/3}$Nb$_{2/3}$O$_3$ (PMN) allowed for the reconstruction  of the
nonequilibrium spin glass state order parameter and its comparison
with the results of recent NMR experiment ( Blinc et al.,
Phys. Rev. Lett {\bf 83}, No. 2 (1999)).
It is shown  that  the degree of the local freezing is rather small 
even at temperatures where  the field-cooled permittivity 
exceeds   the frequency dependent permittivity by an order of 
magnitude. This observation indicates the significant role of the critical 
slowing down (accompanying the glass freezing) in  the system dynamics.  
Also the theory predicts an important interrelationship between the
frequency dependent permittivity and the zero-field-cooled
permittivity, which proved to be consistent with the  experiment
in PMN (A. Levstik et. al., Phys. Rev. B {\bf 57} 11204 (1998)).

\end{abstract}
]
\narrowtext

~~~

\section{introduction}

Relaxor ferroelectrics which are disordered perovskite ferroelectrics
like PbMg$_{1/3}$Nb$_{2/3}$0$_3$ (PMN) or  PbSc$_{1/2}$Ta$_{1/2}$O$_3$
(PST) represent a new class of materials 
which have been a subject of numerous
investigations (see, for example, Refs.\cite{Cross,Burns,Vieland,Mathan,Kleemann,Sommer,VugT,Samara,Glazounov,Egami,Vakhrushev,VugR,Pirc}).
Relaxor ferroelectrics manifest themselves  in the extraordinary low
frequency dispersion of their dielectric permittivity compared with regular 
ferroelectrics. The position and the height of the permittivity maximum 
plotted as a function of temperature depends on the frequency of the
probe field and shifts to lower temperatures when the frequency decreases.

The latter behavior is accompanied by the observed splitting between the 
field-cooled (FC) and zero-field-cooled (ZFC) permittivity and the 
existence of long lived remanent polarization. However, as has been emphasized
recently, the bulk of the relaxation spectra in  relaxors remains
active even far
below the temperature $T_f$ where the FC and ZFC permittivity split, and 
the nonlinear susceptibility does not diverge at $T_f$.   
All these findings indicate  nonequilibrium phenomena and 
quasi-nonergodicity (``freezing''), rather than a true thermodynamic dipole 
spin glass transition.    

On the other hand, the existence of the very high dielectric constant
indicates that these systems are close to ferroelectric instability,
and, therefore, one could expect the manifestation of the critical 
slowing down of dynamics and a  competition between the 
critical slowing down and the dipole spin glass  freezing.
Such a dichotomy makes it a non-trivial task to extract from the
experiment quantitative characteristics of the freezing. 

One of the important characteristics of spin glass freezing is a
value of the Edwards-Anderson order parameter $q$, which in relaxors is a time
dependent quantity. We  

~

~~~~

~~~~~~

\noindent
will show that this quantity can be extracted
from the experimental results on the frequency dependent permittivity,
with the use of the model  discussed below.
Also the theory predicts an important  interrelationship between the
frequency dependent permittivity and the zero-field-cooled
permittivity, which proved to be consistent with the experiment.

\section{model}
It has been recently proposed\cite{VugR} that relaxor behavior is a common
characteristic of the collective dynamics of localized giant  dipole moments
distributed in highly polarizable crystals, and a necessary condition
of relaxor behavior is the simultaneous  existence    
of the broad distribution of the local field and the 
broad distribution of dipole relaxation frequencies.
Also experiments\cite{Egami,Vakhrushev,Mathan} favor the hypothesis
that the physical origin of localized dipole moments is the off-center
shift of the 
atom even at high temperatures. In this model the large values
of the dipole moments are associated with the 
polarization cloud (cluster) formed by the simultaneous displacements
of the other atoms adjacent  to a given off-center ion. 

A convenient approach to describe dynamical  behavior of relaxor 
ferroelectrics is to start from the Bloch type equations widely 
explored in the theory of regular ferroelectrics possessing Debye 
relaxation. We write the Bloch equation in the form 

\begin{equation}
{\partial P_{cl} \over \partial t} ={1 \over \tau} (P_{cl}-
P_{cl}^{eq}(E)).
\label{bloch}
\end{equation}
Eq.(\ref{bloch}) describes the relaxation of polarization of each 
cluster to its quasiequilibrium value $P_{cl}^{eq}(E)$ which depends 
on the value of the local field $E$ induced by other clusters at any
moment of time. In general the local field $E$ is a time dependent
random field. It includes also the contribution from the applied field 
$E_{ex}$ and the contribution from the static random field caused by
material imperfections. Note that  Eq.(\ref{bloch}) is
quite general and, although the explicit form of $P_{cl}^{eq}(E)$ and
the precise definition of the polar clusters are model dependent, it
effects  only the coefficients of the theory discussed below.

In order to apply Eq.(\ref{bloch}) to relaxor ferroelectrics one should
perform an average over the distribution of relaxation times $\tau$
and the random local field $E$.
We assume that the distribution function $f(E,P)$ of the
local field, which depends parametrically on the average polarization
of the system $P(t)$, has the form $f(E,P) = \tilde{f}(E-\gamma P -
\gamma \epsilon_0 E_{ex}/4 \pi)$.  This form of $f(E)$ is consistent
with the mean field approximation $f(E,P)= \delta(E-\gamma P -
\gamma \epsilon_0 E_{ex}/4 \pi)$, where $\delta$ is the Dirac $\delta$-
function and $\gamma$ is the local field phenomenological parameter.
The effect of the local field fluctuations corresponds to the replacement 
of the $\delta$-function by the function $\tilde{f}$ with  finite
width. The value $\gamma \epsilon_0 E_{ex}/4 \pi$ is the local field 
induced by the external  field at the location of each off-center
ion in a dielectric media with the
dielectric constant $\epsilon_0>>1$, i.e., we assume that the
localized dipole moments  are distributed in  crystals with large
lattice permittivity caused by the existence of the soft modes.
This assumption is consistent with the observation in relaxors of the soft
optical modes (remaining finite at all temperatures), frequencies of which 
are of the same order of magnitude ($\sim$ 50cm$^{-1}$) as those in 
highly polarizable dielectrics such as KTaO$_3$, SrTiO$_3$.
 
It is known that the existence of the broad distribution of relaxation 
times leads to  non-exponential behavior  in the polarization relaxation.
In order to reproduce this effect within the proposed formalism we
rewrite Eq.(\ref{bloch}) in the
integral form and then take the average with respect to $\tau, E$, and the 
initial cluster polarization $P_{cl}(0)$. Thus, we obtain  

\begin{eqnarray}
&P(t) =&P(0)q(t)-k(T) \int_0^t dt'{dq(t') \over dt'} [ P(t-t')
\nonumber\\
&     &+{\epsilon_0 \over 4 \pi} E_{ex}(t-t')].
\label{P}
\end{eqnarray}
In Eq.(\ref{P}) we assumed a linear response of the polarization to
the  
applied electric field, and $T>T_c$ where $T_c$ is the temperature
of a possible ferroelectric phase transition,   

\begin{equation}
k(T) = \int dE P_{cl}^{eq}(E) {\partial f(E,P) \over \partial P}|_{P=0}~,
\end{equation}
and for  second order phase transitions 
$k\rightarrow 1$ as $T \rightarrow T_c$.   
The function $q(t)$  is equal to 
\begin{equation}
q(t) =   \int_{\tau_0}^{\tau_m} d \tau \tilde{g}(\tau)e^{-t/\tau}  \approx 
\int_{t}^{\tau_m} d \tau \tilde{g}(\tau) 
\label{q}
\end{equation} 
where $\tilde{g}(\tau)$ is a distribution function of relaxation times and the
right hand side expression of Eq.(\ref{q}) is valid for   smooth
functions $\tilde{g}(\tau)$. 
The variable $q(t)$ describes the fraction of clusters effectively frozen 
at time $t$
and therefore has the  meaning of the Edwards-Anderson  spin glass order 
parameter on a finite time scale. Note that, as we assumed above, 
$q(t) \rightarrow 0$ at $t \rightarrow \infty$.

\section{frequency dependent permittivity} 

The steady state frequency 
dependent permittivity can be easily obtained
from Eq.(\ref{P}) assuming $E_{ex}(t)=E_{ex}^{(1)} e^{i\omega t}$ and using
the definition 

\begin{equation}
\epsilon(\omega,T) =4 \pi {\partial P(\omega) \over \partial E_{ex}^{(1)}}
 +\epsilon_0.
\label{epsilonP}
\end{equation}
Thus we obtain

\begin{equation}
\epsilon(\omega,T) = {\epsilon_0 \over 1- k(T) Q(\omega,T)}
\label{epsilon}
\end{equation}
where 

\begin{equation}
Q(\omega,T)=\int_{\tau_0}^{\tau_m} d\tau e^{i\omega t} {dq \over dt}  
\end{equation}
For the smooth function $\tilde{g}(\tau)$ the real part of $Q(\omega,T)$ can be 
simplified as\cite{Courtens}

\begin{eqnarray}
&Q'(\omega,T)= &\int_{\tau_0}^{\tau_m} dt { \tilde{g}(\tau) \over 
1+ \omega^2 \tau^2}
\approx   \int_{\tau_0}^{1/\omega}d\tau \tilde{g}(\tau)\nonumber \\
&  &=1- q(\omega^{-1},T).
\label{Q}
\end{eqnarray}
The relation(\ref{Q})  between the functions $q(t)$ 
and $Q'(\omega)$ will be employed below to obtained  information on the
 degree of local freezing  in relaxors from the frequency
dependent dielectric constant using the fact that usually $Q''<Q'$.

At $\omega=0$ Eq.(\ref{epsilon}) defines the static or field-cooled 
permittivity 

\begin{equation}
\epsilon_{FC} = {\epsilon_0 \over 1-k(T)}.
\label{epsilonFC}
\end{equation}
On the other hand, at $ \omega \tau_{m} >> 1$ the function $ Q(\omega,T) 
\rightarrow 0$ and therefore $\epsilon(\omega) \rightarrow \epsilon_0$. 
Thus, in our 
model $\epsilon_0$ can be regarded as a high frequency permittivity
$\epsilon_{\infty}$.
Combining equations(\ref{epsilon}) and (\ref{epsilonFC}) we obtain

\begin{equation}
Q(\omega)= {\epsilon(\omega) -\epsilon_0 \over 
   \epsilon_{FC} -  \epsilon_0} \cdot {\epsilon_{FC}\over \epsilon_0}  
\label{QQ}
\end{equation}

Note that in a number of recent publications on 
relaxors\cite{Tagantsev,Kleemann97,Pirc} the analysis of the 
relaxation spectrum $Q(\omega)$ was based on a different relation 
between $Q(\omega)$ and the permittivity, namely 
 
\begin{equation}
Q(\omega)= {\epsilon(\omega) -\epsilon_0 \over 
   \epsilon_{FC} -  \epsilon_0}   
\label{QPirc}
\end{equation}
where we substituted $\epsilon_0$ for $\epsilon_{\infty}$. 
Equation (\ref{QPirc}) was introduced earlier in spin glasses,
assuming that  the relaxation time $\tau$ in Eq.(\ref{Q}) is a 
characteristic of independent cluster-relaxators.
However this equation has a different meaning when applied to  
 relaxor ferroelectrics. Indeed, Eq.(\ref{QPirc}) can be obtained
from Eq.(\ref{bloch}) if we first perform the average over the
distribution of the local random fields at a constant value of $\tau$
and then perform the average over $\tau$. With such a
two-step averaging we arrive at Eq.(\ref{QPirc}) where $\tau$ is
replaced  by $\tau^* =\tau \epsilon/ \epsilon_0$. Note that 
$\tau^*$ is a relaxation time of the collective polar mode with the
wave vector $q \rightarrow 0$,      
undergoing the critical slowing down of dynamics, 
rather than the relaxation time of the  
individual dipoles or clusters. 
Thus the applicability of
Eqs.(\ref{QPirc}), and  (\ref{Q}) to a  system with long range or 
mesoscopic polar order implies that the 
crystal can be divided into macroregions within  which the relaxation
time of
all dipole moments has the same magnitude, and the average over $\tau$
means the average of the dielectric response for different macroregions.
On the other hand, Eq.(\ref{QQ})  is consistent with the formation of
the short range clusters in  the absence of correlations between the  
relaxation times of  different clusters.

\section{zero-field-cooled susceptibility and remanent polarization}

The values of  $\epsilon_{ZFC}(t)$ can be obtained by solving 
Eq.(\ref{P}) with $P(0)=0$ and $E_{ex}=const$. We consider here a
particular case when one can approximately neglect the memory effects 
in Eq.(2) by replacing $P(t-t') \approx P(t)$ ( which implies a fast 
decay of $dq/dt$). Thus, we obtain

\begin{equation}
\epsilon_{ZFC}(t) = {\epsilon_0 \over 1-k(1-q(t))}
\label{ZFC}
\end{equation}
One can see that   $\epsilon_{ZFC}(t)$  is identical to   $\epsilon(\omega)$
given by Eq.(\ref{epsilon}), 
if one substitutes $\omega$ by $1/t$  in Eq.(\ref{Q}).

In order  obtain the remanent polarization $P_r(t)$ we 
solve Eq.(\ref{P}) in the same manner as above, but with $E_{ex} =0$
and $P(0) \neq 0$. We obtain 
 
\begin{equation}
{P(t) \over P(0)} = {q(t) \over 1-k(1-q(t))} =
{\epsilon_{FC}- \epsilon_{ZFC} \over \epsilon_{FC} - \epsilon_0}
\label{Pr}
\end{equation}
The denominator in Eq.(\ref{Pr}) reproduces the effect of the critical
slowing down of dynamics which is imposed on the effect of glass
freezing characterized by the slow decaying function $q(t)$.
For example, when $k \rightarrow 1 $ ( {\it i.e., } the system is 
in the vicinity of the second order phase transition temperature )
it follows from Eq.(\ref{Pr}) that $P_r(t) = const $ independently of
the value of $q(t)$.

\section{COMPARISON WITH THE EXPERIMENT in PMN}

\subsection{Nonequilibrium spin glass order parameter}

The nonequilibrium spin glass order parameter $q(t,T)$ is a very important
quantity determining the dielectric response of relaxors. The
dielectric  permittivity   can be formulated in terms of 
the  parameter $q(t,T)$ in a way consistent 
with the description of magnetic susceptibility in magnetic alloys 
where  spin glass and ferromagnetic order  coexist. 
It is convenient to  rewrite  Eq.(\ref{epsilon}) in the identical form 

\begin{equation}
\epsilon(\omega,T) = {k \epsilon_0 (1-q) \over 1-k(1-q) } +\epsilon_0,
\label{epsilonSK}
\end{equation}
which separates the contribution to the permittivity from  the
localized dipole moments and the lattice permittivity $\epsilon_0$.
The first term in Eq.(\ref{epsilonSK}) is very similar to the well
known solution  for the susceptibility given by the infinite range 
Sherrington-Kirkpatrick model (mean field theory)  

\begin{equation}
\chi= {C (1-q) \over T-\theta(1-q)}
\label{SK}
\end{equation}
Equation (\ref{SK})  has been widely used  for 
the description of the experiments on ZFC susceptibility as well as
frequency dependent susceptibility  
in magnetic alloys\cite{Binder} and relaxors\cite{Glinchuk} by
treating the parameters $C$ and $\theta$ as 
purely phenomenological fitting parameters. In fact, our approach justifies  
and generalizes Eq.(\ref{SK}) in the case of nonequilibrium spin glass state.  
The deviation of the parameter $k(T)$ from $\theta /T$ indicates the 
deviation from the mean field picture. 

Equations (\ref{epsilon}) and (\ref{Q}) allow  reconstructing the
values of the 
functions $q(t,T)$ from  experiments on the frequency dependent permittivity. 
 We applied them to  the recent data on PMN  by Levstic {\it et
al}\cite{Pirc}  who
obtained the values  of the frequency dependent permittivity, as well
as field-cooled and
zero-field-cooled permittivity.

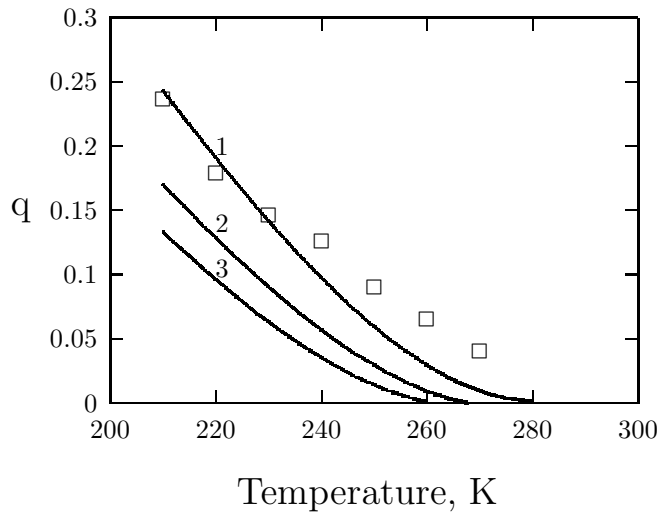
\begin{figure}
\hskip -0.7cm
\input{pmn_q_fig.tex}
\caption{ Reconstructed values of the nonequilibrium spin glass order
parameter in PMN as a function of temperature corresponding to the following  
frequencies of the applied field: 100kHz (1), 1kHz (2), and 20Hz (3).
Squares reproduces the data obtained in Ref.[19].}
\end{figure}

We used $\epsilon_0 \approx 1000$ corresponding to the  values of $\epsilon$ 
in the low temperature limit. The results of the reconstruction are presented  
in Fig.1 where  the temperature dependences of  $q(T)$ as shown for three 
distinct times corresponding to the  frequencies of the 
applied field used in the experiment equal to  20Hz, 1kHz, and 100kHz. 
One can see that the values of the $q(T)$ are rather small even at
T=200K where 
the values of $\epsilon(\omega)$ are almost 10 times less than the
values of  
$\epsilon_{FC}$. The explanation of this important effect lies in the 
extremely high  dielectric constant of PMN leading to  $k \approx
0.98$ at $T$ =200K. Thus, the system is extremely close to 
the ferroelectric instability which results in the significant
effect of the critical slowing down, accompanying the glass freezing, 
upon the system dynamics.

In Fig.1 the reconstructed values of $q$ are compared with the the
data obtained from the recent experiments by Blinc {\it at al.}\cite{Blinc}
on the observation of inhomogeneous broadening of NMR lines of Nb ions.
The appearance of inhomogeneous broadening signals the slowing down
of the dynamics on the scale of inhomogeneous linewidth (equal 
to 10-50 kHz).    
A reasonable correspondence  between  curve 1 and 
the data\cite{Blinc} (with the account of comparable time scales in both 
experiments) 
allows us to suggest the following scenario for the NMR broadening
mechanism in PMN. As known, e.g., from the analysis of nuclear spin 
lattice relaxation of 
Li and Nb ions in KTaO$_3$:Li, Nb\cite{VugG}, the appearance of 
polarization on the mesoscopic length scale in highly polarizable
crystals results in a 
significant modulation of the electric field gradient which leads to
the modulation  of nuclear resonance frequency. Since in disordered PMN 
crystals all atoms lack a center of symmetry, one can assume the
following form of the nuclear frequency shift of Nb atoms 

\begin{equation}  
 \Delta\nu = \alpha P_{cl}
\label{nu}
\end{equation}
where $\Delta\nu$ is the frequency shift of the atoms belong to
the given polar cluster.
The width $\delta$ of the inhomogeneous  line will be
proportional to the fraction of polar clusters ( given by the
nonequilibrium spin glass parameter $q(T)$)  whose dymamics is
effectively frozen on the time scale of $1/\delta$. 
The above broadening mechanism gives also a natural explanation of the 
Gaussian line shape observed in Ref.\cite{Blinc} since the
distribution function of cluster polarization is  close to Gaussian in 
highly polarizable crystals\cite{VugG}, 
and, according to Eq.(\ref{nu}), the shape of 
inhomogeneous  NMR line coincides with the shape of the
distribution function of cluster polarization.

\subsection{Interrelationship between ZFC and frequency dependent permittivity}

Another practical aspect of the proposed model, which  helps to 
clarify the  experimental situation in PMN, is the  interrelationship 
between the frequency dependent permittivity and zero-field-cooled
permittivity obtained in Sec.4. 
In order to employ this interrelation note first that in PMN, at the
temperatures nearby  the
temperature of the permittivity maximum, the characteristic relaxation
time satisfies the Vogel-Fulcher  (FV) law\cite{Vieland,Pirc}.   
It allows to replace $\tau$ in Eq.(\ref{q}) by 
\begin{equation}
\tau= \tau_0 e^{U \over T-T_0}  
\end{equation}
and introduce the temperature independent distribution function $g(U)$
of potential barriers\cite{Binder,Courtens}. We obtain ($t > \tau_0$)  
\begin{equation}
q(t) = 1-\int_0^{(T-T_0) \ln ({t \over \tau_0})} d U g(U)
\end{equation}
Thus, $q$ is a function of $(T-T_0)\ln(t/\tau_0)$ that leads to the validity 
of the scaling relation $q(t,T)=q(t_1,T_1)$ with 
\begin{equation}
T_1= (T-T_0) {\ln({t \over \tau_0}) \over  \ln({t_1\over \tau_0})} +T_0
\label{T1}
\end{equation}
Using Eqs.(\ref{T1}), (\ref{epsilon}), and (\ref{Q})
one can reconstruct the values of $q(t, T)$ and $Q'(\omega,T)$ and, therefore,
the values of $\epsilon(\omega)$ by employing the experimental data on 
$\epsilon(\omega_1)$  at a  given frequency $\omega_1$.

The reconstructed values of $\epsilon(\omega)$ at $\omega =20Hz$ are 
presented in Fig.2 (solid line)  where the data at  $\omega_1 =1kHz$
have been taken as a reference.  The parameters of the fit are $T_0
\approx 223$K and $\tau_0 \approx 4 \cdot 10^{-11} sec $, which
are almost the same as obtained in Ref.\cite{Pirc}  $T_0 \approx 224$K
 and $\tau_0 \approx 4.3 \cdot 10^{-11} sec$.  One can see that at high 
temperatures $(T> 245K)$ the reconstruction reproduces the experimental data 
with good accuracy.
The deviation between the reconstructed and experimental data for
lower temperatures is due to violation of the VF law in PMN already discussed
in Refs.\cite{Vieland,Pirc}.       
The self-consistency of the model has been tested also by reconstructing 
the values of the zero-field cooled permittivity using  Eq.(\ref{ZFC})
and the  same reference data as above.  In order to find the time $t$
entering  Eq.(\ref{ZFC}) we use the fact that in the
experiment\cite{Pirc} the dielectric constant was measured  
by slowly  heating the crystal from 80K at the rate 0.5 K/min. 
It gives e.g., the estimate $t$ =330 min at $T=245$K.  
The result of the reconstruction is shown in Fig.2 by the solid line.
As one can see from Fig.2,  the position and the hight of the maximum of 
 $\epsilon_{ZFC}$ are reproduced with very good accuracy. 
The  deviation between the reconstructed values and  experimental data 
is due to violation of the VF law taking place at low temperatures.
Indeed, the  temperature of the 
maximum of  $\epsilon_{ZFC}$ corresponds to $T\approx 245$K,
which, as discussed above, is the boundary value for the  validity of
the VF law.
If the VF law would be valid for all temperatures then  both  $\epsilon_{ZFC}$
and $\epsilon(\omega)$ would approach the ~ value $\epsilon_0$~ at~ $T=T_0$.
One can see
\begin{figure}
\hskip -0.7cm
\input{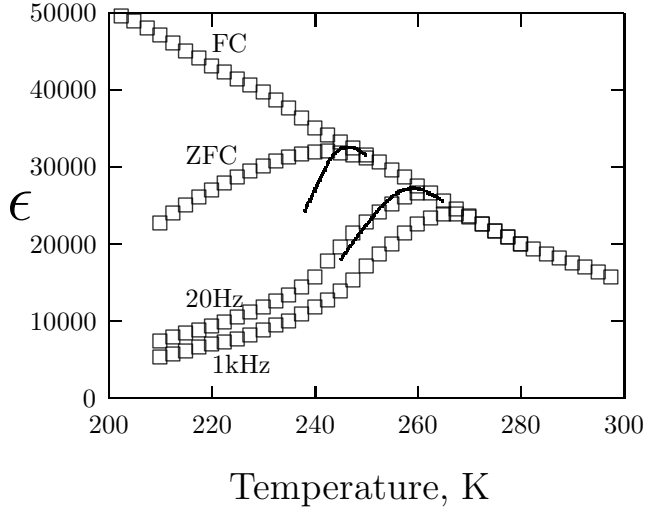}
\caption{Experimental (squares) and reconstructed (solid lines) values
of the dielectric permittivity in PMN. Experimental values are
reproduced from Ref.[13]}  
\end{figure} 

\noindent
 that  this tendency is reproduced by the interpolation of the
solid curves  in Fig.2 to lower temperatures, thus clarifying  the
origin of the observed 
low temperature deviations between the reconstructed and experimental
values  for $\epsilon_{ZFC}$. 

\section{Conclusion}

The high dielectric constant of relaxor ferroelectrics signals that these
materials are close to ferroelectric instability which  manifests itself in
the critical slowing down of dynamics. In disordered relaxor materials 
the critical slowing down  is accompanied by glass like freezing or
cluster dynamics, which could be characterized by the non-equilibrium 
spin glass order parameter. 
We have developed a dynamical model of the dielectric response in
relaxors which explicitely takes into account  the coexistence of
critical and cluster dynamics and allows for separating  them from
the experimental observation.

\acknowledgements
We acknowledge the  support  by the National Science Foundation
and the Princeton Material Institute.

\end{document}